\documentclass[12pt]{article}

\usepackage{amsmath}

\usepackage{times}
\usepackage{bm}
\usepackage{xcolor}
\usepackage[textsize=tiny, colorinlistoftodos]{todonotes}  
\usepackage{comment} 

\usepackage[utf8]{inputenc}
\usepackage[english]{babel}
\usepackage{amsfonts}
\usepackage{amssymb}
\usepackage{color}
\usepackage{colordvi}
\usepackage{lscape}
\usepackage{rotating}
\usepackage{graphicx}
\usepackage{pgfplots}
\usepackage{tikz}
\usepackage[round]{natbib}
\usepackage{algorithm}
\usepackage{animate}
\usepackage{threeparttable}

\pgfplotsset{compat=newest}
\pgfplotsset{plot coordinates/math parser=false}

\DeclareMathOperator*{\plim}{plim}

\makeatletter
\def\BState{\State\hskip-\ALG@thistlm}
\makeatother
\def\1{1\!{\rm l}}

\newlength\figureheight
\newlength\figurewidth

\usepackage{enumitem,float,rotating,multirow,algorithm,algpseudocode,graphicx,lscape,tabularx,ragged2e}

\newtheorem{theorem}{Theorem}
\newtheorem{assumption}{Assumption}

\newtheorem{corollary}{Corollary}

\newtheorem{proof}{Proof}

\usepackage[colorlinks=true, citecolor=blue, linkcolor=blue]{hyperref}

\usepackage{graphicx,bm}

\title{Testing model specification in approximate Bayesian computation\thanks{We acknowledge supercomputing resources made available by the Centro de Computación Científica Apolo at Universidad EAFIT (http://www.eafit.edu.co/apolo) to conduct the research reported in this scientific product. All errors are our own.}}

\author{Andr\'es Ram\'irez-Hassan\thanks{Department of Economics, Universidad EAFIT, Medell\'in, Colombia; email: aramir21@eafit.edu.co}\and David T. Frazier\thanks{Department of Econometrics and Business Statistics, Monash University, Melbourne, Australia; email: david.frazier@monash.edu}}

\date{\today}

\begin{document}
	\maketitle \thispagestyle{empty}
	
	\begin{abstract}
		\setlength{\baselineskip}{10pt}
		\vspace{0.2in} \noindent We present a procedure to diagnose model misspecification in situations where inference is performed using approximate Bayesian computation. We demonstrate theoretically, and empirically that this procedure can consistently detect the presence of model misspecification. Our examples demonstrates that this approach delivers good finite-sample performance and is computational less onerous than existing approaches, all of which require re-running the inference algorithm. An empirical application to modelling exchange rate log returns using a g-and-k distribution completes the paper.  
		
		\vspace{0.2in} \noindent
		
	\end{abstract}
	\newpage
	\setcounter{page}{1}
	\pagenumbering{arabic}
	
	\section{Introduction}
	Approximate Bayesian computation is a likelihood-free inference procedure that allows researchers to conduct Bayesian inference in settings where the likelihood is intractable or otherwise computational demanding. While initially introduced in the population genetics literature, see, for instance, \cite{tavare1997inferring} and \cite{Beaumont2025}, approximate Bayesian computation is now applied in a wide-range of fields including: biology, cosmology and economics, among others; see \cite{sisson2018handbook} for a handbook treatment.
	
	Under correct model specification, \cite{li2018asymptotic} and \cite{frazier2018asymptotic} demonstrate that approximate Bayesian computation displays favourable asymptotic properties, and is asymptotically equivalent to likelihood-based Bayesian procedures that condition on a set of summaries rather than the full dataset. However, \cite{frazier2020model} demonstrate that if the model assumed to have generated the data is misspecified, then the asymptotic behave of approximate Bayesian computation can degrade rapidly. While the approximate Bayesian computation posterior maintains concentration onto some well-defined point in the parameter space when the model is misspecified, the posterior shape is asymptotically non-Gaussian, and the behavior of the posterior mean is unknown in general. Furthermore, in misspecified models, the behavior of commonly applied regression adjustment approaches can produce posteriors that differ markedly from their simpler accept/reject counterparts, with the specific behavior of the posterior depending on the intricacies of the adjustment approach, and have little to do with the actual problem at hand. 
	
	Common approaches to asses model specification in approximate Bayesian computation have so far been based on predictive p-values \citep{bertorelle2010abc}, simulated goodness of fit statistics \citep{bertorelle2010abc} and \citep{lintusaari2017fundamentals}, and discrepancy diagnostics \citep{frazier2020model}. Obtaining the distribution for the latter two statistics under correct model specification requires re-running, many times, the approximate Bayesian computation algorithm: draws from the original posterior distribution are used to generate many pseudo-observed data sets; for each data set the inference algorithm is re-implemented and used to calculate a discrepancy measure; and the distribution of this discrepancy measure under correct specification is then taken as the Monte Carlo distribution of the discrepancy measure over the pseudo-observed data sets, and the relized value of the discrepancy measure compared against this distribution. 
	
	In contrast to existing computationally onerous procedures, we propose a simple statistic to test model misspecification. We demonstrate that this statistic reliably detects departures from correct model specification, and does not require re-running any inference algorithm. We demonstrate that hypothesis tests based on this statistic have asymptotically correct size when the model is correctly specified, and can consistently detect model misspecification. 
	
	We compare this new procedure against existing diagnostic approaches, and find that our approach performs similarly to existing approaches in simple examples, such as a misspecified normal example. However, in realistic data scenarios we find that our testing approach outperforms existing approaches. Across all examples, we find that our approach is much less computationally demanding than existing approaches. 
	
	\section{Model misspecification in approximate Bayesian computation}\label{sec: ABC}
	We assume the observed data ${y}=(y_{1},...,y_{n})^{\top}$ is generated from a complex class of parametric models $\{\theta\in\Theta:P_{\theta}\}$, that depends on unknown parameters $\Theta\subseteq\mathbb{R}^{k_{{\theta}}}$, for which our prior beliefs over $\Theta$ are represented by the density $\pi(\theta)$. Using the observed data $y$, the model $P_\theta$, and the prior density $\pi(\theta)$, Bayes Theorem delivers to us the cornerstone of Bayesian statistics: the posterior density $\pi({\theta \mid y})\propto p(y\mid\theta)\pi({	\theta })$. 
	
	Generally, exact Bayesian inference requires that  $\pi({\theta \mid y})$ be available
	in closed-form, at least up to the constant of
	proportionality, which ultimately entails that the model density $p(y\mid\theta)$ be tractable. On the other hand, approximate Bayesian inference schemes generally remain applicable in cases where $p(y\mid {\theta})$ is computationally too costly to compute or not accessible via numerical algorithms; see the unpublished 2021 technical report by Martin, Frazier and Robert for a review of approximate Bayesian methods more broadly (https://arxiv.org/abs/2112.10342). 
	
	As noted, approximate Bayesian computation has received significant attention due to the ease with which it can be implemented even in very complex models: so long as we can simulate from the assumed model, we can conduct inference on the model unknowns. The aim of approximate Bayesian computation is to build a reliable approximation to $\pi({\theta \mid y})$ in cases where $p(y\mid \theta)$ is not accessible. Approximate Bayesian computation is predicated on the belief that the observed data ${y}$ is drawn from one of the constituent members in the class $\{{\theta \in \Theta }:P_{{\theta }}\}$, and inference on the unknown $\theta$ is conducted by first drawing $\theta\sim\pi(\theta)$, simulating pseudo-data ${z}$, ${z}=(z_{1},...,z_{n})^{\top}\sim P_{\theta}$, and ``comparing'' ${z}$ with the observed data ${y}$. In most cases, this comparison is carried out using a vector of summary statistics $\eta(z)\in\mathcal{B}\subseteq\mathbb{R}^{k_\eta}$, with $k_\eta>k_\theta$, and a norm $d\{\cdot,\cdot\}$. 
	
	Simulated values of $\theta$ are then accepted, and used to build an approximation to the exact posterior, if the distance $d\{\eta(z),\eta(y)\}$ is small relative to a pre-defined tolerance parameter $\epsilon$, which shrinks to zero as the sample size $n$ goes to infinity. The most basic form of the approximate Bayesian computation algorithm is the accept/reject algorithm.
	\vspace{-.25cm}
	\begin{algorithm}
		\caption{Accept and reject: Approximate Bayesian computation}\label{ABC0}
		\begin{tabbing}
			\enspace {(1) For \texttt{$i=1,\dots,N$} do}\\
			\qquad (i) Simulate ${ {\theta} }^{i}$ from $\pi({ {\theta} })$\\
			\qquad (ii) Simulate ${ z}^{i}=(z_{1}^{i},...,z_{n}^{i})^{\top}$ from the model, $p(\cdot\mid{{\theta} }^{i})$\\
			\qquad (iii) Calculate $\epsilon_{(i)}=d\{{\eta }({ y}),{ \eta }({ z}^{i})\}$\\
			\enspace (2) Order the distances $\epsilon_{(1)}\leq\cdots\leq \epsilon_{(N)}$\\
			\enspace (3) Select all ${\theta}^i$ associated with the $\alpha=\delta/N$ smallest distances, with $\delta$ much\\
			smaller than $N$.\\		 
		\end{tabbing}
		\vspace*{-25pt}
	\end{algorithm}

	It can be shown that Algorithm \ref{ABC0} produces draws from the following approximation to $\pi(\theta\mid y)$: 
	\begin{flalign}
		\pi_{\epsilon }\{{\theta \mid \eta (y)}\}&=\int_{\mathcal{Z}}\pi_{\epsilon}\{{\theta,z \mid\eta (y)}\}dx z=\frac{P_{\theta}\left[d\{\eta(z),\eta(y)\}\leq \epsilon\right]dx \Pi(\theta)}{\int_\Theta P_{\theta}\left[d\{\eta(z ),\eta(y)\}\leq \epsilon\right]dx \Pi(\theta)}\label{ABC_post},
	\end{flalign}
	where 
	$P_{\theta}\left[d\{\eta(z),\eta(y)\}\leq \epsilon\right]=\int_{\mathcal{Z}}\1\left\{d\{\eta(z),\eta(y)\}\leq \epsilon\right\}p(z\mid \theta)dx z $. We note that the posterior $\pi_{\epsilon }\{{\theta \mid \eta (y)}\}$ depends explicitly on the chosen tolerance $\epsilon$, and is conditioned on $\eta(y)$, not the full data set $y$.  
	
	It is well-known that Algorithm \ref{ABC0} is inefficient since all draws are generated independently from the prior. More efficient sampling schemes can also be implemented by replacing the accept/reject step with more informative proposals \citep{marin2012approximate, sisson2007sequential,Beaumont2009adaptivity}; see chapter 4 of \cite{sisson2018handbook} for a review. While these methods allow us to obtain more readily accessible accepted draws, each accepted draw is still from the posterior in \eqref{ABC_post}. Consequently, in terms of examining model misspecification, the simplest accept/reject algorithm with suffice for our purposes. 
	
	Under correct specification of the assumed model, \cite{frazier2018asymptotic} show that so long as $\alpha_n=\delta/N_n$ in Algorithm \ref{ABC0} converges to zero fast enough, as the sample size $n$ diverges, the approximate Bayesian computation posterior $\pi_{\epsilon}\{\theta\mid \eta( y)\}$ concentrates onto values of $\theta$ such that $b_0=\plim_n \eta(y)$ and $b(\theta)=\plim_n\eta(z)$ agree. Throughout, we let $\plim_n X_n$ denote the probability limit of the random sequence $\{X_n:n\ge1\}$ as $n\rightarrow+\infty$, if it exists.  
	
	However, if the model $P_{{\theta}}$ used to simulate pseudo-data $z$ is misspecified, in the sense that there is no $\theta\in\Theta$ such that $b_0=b(\theta)$, then the model is said to be misspecified \citep{frazier2020model}. Model misspecification creates at least two practical issues for inference on ${\theta}$ via Algorithm \ref{ABC0}. As shown in \cite{frazier2020model}, when the model is misspecified the posterior displays non-standard asymptotic behavior, does not produce reliable credible sets, and the behavior of its posterior moments are suspect. In addition, when the model is misspecified, commonly applied regression post-processing approaches, like linear regression adjustment \citep{beaumont2002approximate}, produce a posterior that differs substantially from $\pi_{\epsilon}\{\theta\mid \eta( y)\}$ and whose posterior moments can be arbitrarily far away from those of $\pi_{\epsilon}\{\theta\mid \eta( y)\}$ \citep{frazier2020model} (please see the Appendix for further details).
	
	\section{A test for model misspecification}\label{sec: test}
	Given the prominence of model misspecification in empirical applications, and the potential issues that this can cause when conducting approximate Bayesian computation, it is paramount for researchers to be able to determine whether or not $P_{{\theta}}$ is misspecified.
	
	In this section we provide a mechanism for determining model specification that is based on the asymptotic results in \cite{li2018asymptotic} and \cite{frazier2018asymptotic}. Given $\pi_\epsilon\{\theta\mid\eta(y)\}$, define the posterior mean $\hat{{\theta}}=\int_{ \Theta}{\theta} \pi_{\epsilon}\{{\theta}\mid \eta( y)\}dx\theta$, and consider generating a sequence of pseudo-data sets $\hat{z}^i \  (i=1,\dots,N_n)$, each simulated under $\hat\theta$. For $\hat{V}_0$ denoting any consistent estimator of $\plim_n\text{var}\left[n^{1/2}\{\eta( y)- b_0\}\right]$, we propose researchers use the following statistic to detect model misspecification in simulation-based Bayesian inference:
	$$\mathcal{J}\{\hat{\eta}( z),\eta( y)\}={n}^{1/2}\{\hat{\eta}( z)-\eta( y)\}^{\intercal} \hat{V}_{0}^{-1}{n}^{1/2}\{\hat{\eta}( z)-\eta( y)\}.$$
	
	To ensure that the simulation of pseduo-data under $\hat\theta$ does not impact the behavior of $\mathcal{J}\{\hat{\eta}( z),\eta( y)\}$, the number of Monte Carlo draws $N_n$ should satisfy $N_n\geq C\log(n){n}^{q/2}$ for some $C>0$, and $q=\max\{k_\theta,2\}$. We can then show that the above statistic has useful behavior under the regularity conditions in \cite{frazier2018asymptotic}; we reproduce these regularity conditions in the Appendix; proofs of all stated results are also given in the Appendix. Let $\chi^2_k$ denote a Chi-squared distributed random variable with $k$ degrees of freedom.
	
	\begin{theorem}\label{thm1}Under the conditions of Theorem 2 in \cite{frazier2018asymptotic}, see the Appendix for details, $\mathcal{J}\{\hat{\eta}( z),\eta( y)\}$ converges in distribution to a $\chi^2_{k_\eta-k_\theta}$ random variable.
	\end{theorem}
	
	In the case of independently and identically distributed data with $\eta( y)=\sum_{i=1}^{n} \eta(y_i)/n$, the variance estimator can simply be taken as $\hat{ V}_{0}=n^{-1}\sum_{i=1}^{n}\{ \eta(y_i)-\eta( y)\}\{ \eta(y_i)-\eta( y)\}^{\intercal}$. More generally, we note that the bootstrap may also be used to obtain $\hat{ V}_{0}$. 
	
	The statistic $\mathcal{J}\{\hat{\eta}( z),\eta( y)\}$ can be used to conduct a hypothesis test of  $$\mathrm{H}_{0}:\inf_{{\theta}\in \Theta}d\{ b(\theta), b_0\}=0,\text{ versus } \mathrm{H}_{\mathrm{a}}:\inf_{{\theta}\in \Theta}d\{ b(\theta), b_0\}>0,$$
	
	\noindent where the critical value for the test statistic is obtained from a $\chi^{2}_{k_{\eta}-k_{{\theta}}}(1-\alpha)$ distribution. This test shares similar structure to the indirect model specification test in \cite{gourieroux1993indirect}. Under $\mathrm{H}_{\mathrm{a}}$, we have the following result.
	
	\begin{corollary}\label{cor1}Under $\mathrm{H}_{\text{a}}$, and the regularity conditions in Theorem 2 of Frazier et al. (2018), $\mathrm{pr}\left[\mathcal{J}\{\hat{\eta}( z),\eta( y)\}\ge \chi^2_{k_\eta-k_\theta}(1-\alpha)\right]\rightarrow 1$ as $n\rightarrow\infty$ for $\alpha\in(0,1)$.
	\end{corollary}
	
	\section{Simulation exercises}\label{sec: sim}
	
	\subsection{A simple normal example}
	Consider an artificially simple example where the assumed data generating process is independently and identically distributed $z_j\sim  N({\theta},1)$. However, the true data generating process is independently and identically distributed $y_j\sim  N({\theta},\sigma^2)$ $(j=1,\dots,n)$. When $\sigma^2\neq 1$, the model simulating pseudo-data differs from the model generating observed data. We consider as the basis of our analysis the summary statistics $\eta_{1}({y})=\frac{1}{n}\sum_{j=1}^{n}{y}_{j}$
	and  $\eta_{2}({y})=\frac{1}{n}\sum_{j=1}^{n}\left\{{y}_{j}-\eta_{1}({y})\right\}^{2}$.
	
	We consider the outlined testing procedures across a grid of values for $\sigma$ over $(0.8,1.3)$. For each value of $\sigma$ in this grid we generate observed data $y_{j}\sim N(0,\sigma)$, and calculate $\eta( y)$. For these Monte Carlo experiments, we set $\pi({\theta})$ equal to $\mathcal{U}(-1,1)$ to draw $ z^i$, and consider $d\left\{ {\eta}( y),{\eta}( z^i)\right\}$ in Algorithm \ref{ABC0} to be the Euclidean norm, we take $N$=50,000 Monte Carlo draws and choose $\alpha=0.01 \gtrsim n^{-k_{ \theta}/2}=n^{-1/2}$, $n=(100, 500, 1000)$ \citep{frazier2018asymptotic}. We consider 100 replications using a nominal size equal to 5\%.
	
	We compared the four procedures to detect misspecification. The asymptotic goodness of fit test generating $N_n=10,0000 \geq C\log(n){n}$ pseudo-data $\hat{ z}^i$ to obtain $\hat{\eta}( z)$, and construct $\mathcal{J}\{\hat{\eta}( z),\eta( y)\}$ using $\hat{ V}_0=n \ \text{diag}\left\{\eta_2( y)/n, 2\eta_2( y)^2/(n-1)\right\}$, the predictive p-value diagnostic using $\eta_2( y)=\frac{1}{n}\sum_{i=1}^n(y_i-\bar{y})^2$, the discrepancy diagnostic implementing each approximate Bayesian computation procedure with N = 50,000 and $\alpha=0.01$, using $h=(\theta^2, \theta^3)^{\top}$ \citep{frazier2020model}, and the goodness of fit test based on approximate Bayesian computation re-sampling, the last three tests use $R=100$ replications. See Appendix for details.
	
	\begin{table}[htbp]\centering \caption{Computational average time: Misspecification diagnostics\label{tab:tab1}}
		\begin{threeparttable}
			\resizebox{1\textwidth}{!}{\begin{minipage}{\textwidth}
					\begin{tabular}{c c c c c}
						\multicolumn{5}{c}{A simple normal example$^*$}\\
						
						$n$	&	Asymptotic GoF	&	Simulated GoF	&	Pred. p-value	&	Discrepancy	\\
						100 &	0.46 secs	&	3.46 secs	&	0.09 secs &	2.43 mins	\\
						500	&	0.66 secs	&	3.21 secs	&	0.13 secs 	&	4.88 mins	\\
						1000	&	1.02 secs	&	3.46 secs	&	0.14 secs	& 8.43 mins\\
						\multicolumn{5}{c}{A linear regression using g-and-k distribution with endogeneity$^+$}\\
						500	&	54.96 secs	&	10.45 secs	&	0.29 secs 	&	1.24 mins	\\
						1000	&	1.43 mins &	10.45 secs	&	0.32 secs	& 1.95 mins		\\
						\multicolumn{5}{c}{A nonlinear ecological dynamic system$^+$}\\
						250	&	 12.54 secs	&	1.85 mins	& 0.12 secs 	&	5.57  mins	\\
						500	&	 22.81 secs	&	1.80 mins	&	0.13 secs	& 10.22 mins		\\
						1000	&	  40.87 secs	&	 1.79 mins	&	 0.16 secs	&  19.38 mins
					\end{tabular}
					\begin{tablenotes}[para,flushleft]
						\footnotesize \textit{Notes}: $^*$ Computational average time over 100 data sets and $\sigma$ grid once we have results from Algorithm \ref{ABC0} in an Intel(R) Xeon(R) Gold 6130 CPU @ 2.10GHz 64 GB RAM in parallel computing using 32 cores and R software.\\
						$^+$ Computational average time over 50 data sets, and $\rho$ and $k$ grids (see below) once we have results from Algorithm \ref{ABC0} in an Intel(R) Xeon(R) CPU E5-2670 v3 @ 2.30GHz RAM 397 GB architecture x86\_64 CPU 64-bit in parallel computing using 24 cores and R software.
						
					\end{tablenotes}
			\end{minipage}}
		\end{threeparttable}
	\end{table}
	
	Table \ref{tab:tab1} shows computational average time calculated after implementing Algorithm \ref{ABC0}, which is the point of departure for all four procedures. The predictive p-value algorithm is the least time demanding followed by the asymptotic GoF test. These two diagnostics do not require to re-run approximate Bayesian computation algorithms multiple times. On the other hand, the simulated GoF test is in third position, and the discrepancy diagnostic is by far the most computational demanding. The former requires to re-run Algorithm \ref{ABC0} multiple times, and the latter requires computing both accept and reject and regression approximate Bayesian computation many times.    
	
	Table \ref{tab:tab2} displays the probability of rejecting the null hypothesis of no misspecification. All tests seem to be consistent, but there are some differences regarding the size.
	
	\begin{table}[htbp]\centering \caption{Power results of misspecification tests\label{tab:tab2}}
		\begin{threeparttable}
			\resizebox{0.9\textwidth}{!}{\begin{minipage}{\textwidth}
					\begin{tabular}{c c c c c c}
						\multicolumn{6}{c}{A simple normal example$^*$}\\
						$n$	&	$\sigma$	&	Asymptotic GoF	&	Simulated Gof	&	Pred. P-value	&	Discrepancy	\\
						100	&	0.8	&	0.99	&	0.79	&	0.94	&	0.59	\\
						100	&	0.9	&	0.48	&	0.18	&	0.36	&	0.08	\\
						100	&	1	&	0.09	&	0.05	&	0.08	&	0.01	\\
						100	&	1.1	&	0.15	&	0.31	&	0.33	&	0.22	\\
						100	&	1.2	&	0.65	&	0.80	&	0.82	&	0.69	\\
						100	&	1.3	&	0.89	&	0.95	&	0.96	&	0.92	\\
						500	&	0.8	&	1.00	&	1.00	&	1.00	&	1.00	\\
						500	&	0.9	&	0.96	&	0.90	&	0.94	&	0.85	\\
						500	&	1	&	0.04	&	0.04	&	0.05	&	0.04	\\
						500	&	1.1	&	0.79	&	0.85	&	0.90	&	0.79	\\
						500	&	1.2	&	1.00	&	1.00	&	1.00	&	1.00	\\
						500	&	1.3	&	1.00	&	1.00	&	1.00	&	1.00	\\
						1000	&	0.8	&	1.00	&	1.00	&	1.00	&	1.00	\\
						1000	&	0.9	&	1.00	&	1.00	&	0.99	&	0.99	\\
						1000	&	1	&	0.03	&	0.00	&	0.06	&	0.03	\\
						1000	&	1.1	&	0.99	&	0.99	&	0.99	&	0.97	\\
						1000	&	1.2	&	1.00	&	1.00	&	1.00	&	1.00	\\
						1000	&	1.3	&	1.00	&	1.00	&	1.00	&	1.00	\\
						
						\multicolumn{6}{c}{A linear regression using a g-and-k distribution with endogeneity$^+$}\\
						$n$	&	$\rho$	&	Asymptotic GoF	&	Simulated Gof	&	Pred. P-value	&	Discrepancy	\\
						500	&	0	&	0.14	&	0.00	&	0.00	&	0.00	\\
						500	&	0.4	&	0.24	&	0.00	&	0.00	&	0.00	\\
						500	&	0.8	&	0.98	&	0.00	&	0.00	&	0.00	\\
						1000	&	0	&	0.10	&	0.00	&	0.00	&	0.00	\\
						1000	&	0.4	&	0.70	&	0.00	&	0.00	&	0.00	\\
						1000	&	0.8	&	1.00	&	0.00	&	0.00	&	0.04 	\\
						\multicolumn{6}{c}{A nonlinear ecological dynamic system$^-$}\\
						$n$	&	$k$	&	Asymptotic GoF	&	Simulated Gof	&	Pred. P-value	&	Discrepancy	\\
						250	&	0.6	&	0.48	&	0.00	&	0.00	&	0.02	\\
						250	&	0.7	&	0.43	&	0.00	&	0.00	&	0.06	\\
						250	&	0.8	&	0.22	&	0.00	&	0.00	&	0.04	\\
						250	&	0.9	&	0.24	&	0.00	&	0.00	&	0.04	\\
						250	&	1	&	0.14	&	0.00	&	0.00	&	0.02	\\
						500	&	0.6	&	0.82	&	0.00	&	0.00	&	0.16	\\
						500	&	0.7	&	0.57	&	0.00	&	0.00	&	0.06	\\
						500	&	0.8	&	0.46	&	0.00	&	0.00	&	0.06	\\
						500	&	0.9	&	0.47	&	0.00	&	0.00	&	0.02	\\
						500	&	1	&	0.33	&	0.02	&	0.02	&	0.08	\\
						1000	&	0.6	&	1.00	&	0.00	&	0.00	&	0.10	\\
						1000	&	0.7	&	0.94	&	0.00	&	0.00	&	0.10	\\
						1000	&	0.8	&	0.71	&	0.00	&	0.00	&	0.10	\\
						1000	&	0.9	&	0.53	&	0.00	&	0.00	&	0.10	\\
						1000	&	1	&	0.36	&	0.00	&	0.00	&	0.04
						
					\end{tabular}
					\begin{tablenotes}[para,flushleft]
						\footnotesize \textit{Notes}: Probability of rejecting the null hypothesis of no misspecification. Approximate Bayesian computation assumes $^*\sigma=1$, $^+\rho=0$, and $^-k=1$, and the data generating processes may have other values.
						
					\end{tablenotes}
			\end{minipage}}
		\end{threeparttable}
	\end{table}

	\subsection{A linear regression analysis using a g-and-k distribution with endogeneity}
	
	Quantile distributions do not have a density function in closed form \citep{drovandi2011likelihood}; they are defined through their inverse cumulative distribution functions, $Q(p\mid{\theta})=F^{-1}(p\mid{\theta})$, where $F=P(U\leq u)$, and $Q$ is equal to the $p$-quantile \citep{rayner2002numerical}. 
	
	Here we consider an example where the assumed data generating process is $z_j=x_j\beta+u_j$ $(j=1,\dots,n)$, $u_j$ distributes independently and identically with a g-and-k quantile distribution.
	
	$$Q^{gk}\left\{z(p)\mid{\theta}\right\}=a+b\left[1+c\frac{1-\exp\left\{-gz(p)\right\}}{1+\exp\left\{-gz(p)\right\}}\right]\left\{1+z(p)^2\right\}^k z(p),$$
	
	where $z(p)$ is the standard normal quantile function, and $c=0.8$ is suggested.
	
	On the other hand, we assume that the true data generating process is $y_j=x_j\beta+u_j$ where 
	
	$$v_j=Q^{gk}\left\{z(p)\mid\theta_j\right\}, \  v_j=(x_j,u_j), \ \begin{Bmatrix} z(p_x) \\ z(p_u)\end{Bmatrix}\sim N \begin{Bmatrix} \begin{pmatrix} 0\\ 0\end{pmatrix}, \begin{pmatrix} 1 & \rho \\ \rho & 1\end{pmatrix} \end{Bmatrix}.$$
	
	We took into account a grid of values for $\rho$ over $(0, 0.8)$, $\rho\neq 0$ implies that inference should be based on the joint likelihood function, $y_{j}=0.5x_j+u_j$ where $a_x=a_u=0$, $b_x=b_u=1$, $c_x=c_u=0.8$, $g_x=g_u=2$, $k_x=k_u=1$, and $n=(500, 1000)$ in 50 simulation exercises. We perform inference regarding $\beta$ and $k$ in this example, and fixed $a, b$ and $g$ at their population values. We consider as the basis of our analysis the summary statistics $\eta_{1}({y})=\sum_{i=1}^{n}x_iy_i/\sum_{i=1}^{n}x_i^2$, $\eta_2({y})=L_3-L_1$, $\eta_3({y})=(L_3+L_1-2L_2)/\eta_2$  and $\eta_4({y})=(E_7-E_5+E_3-E_1)/\eta_2$ where $L_l$ is the $l$th quartile and $E_l$ is the $l$th octile of $y_j-x_j\eta_1({y})$ (Drovandi \& Pettitt, 2011).
	
	We set $\pi(\theta)=\pi(\beta,k)$ equal to $\mathcal{U}(0,5)\times \mathcal{U}(0,5)$, consider the Euclidean norm in Algorithm \ref{ABC0}, $N$=100,000 Monte Carlo draws and choose $\alpha=0.001\gtrsim n^{-1}$. We implement Algorithm 1 in Appendix with $N_n=10,000\geq C\log(n){n}$, and calculate $\hat{{V}}_0$ using 200 bootstrap samples. Algorithms 2, 3, and 4 in Appendix use $R=100$, where $\eta_1(\hat{ z}_r)$ is used for the predictive p-value test, and the approximate Bayesian computation in the the discrepancy diagnostic test use $N=10,000$ and $\alpha=0.01$ with $h =(\theta^{2\top} \ \theta^{3\top})^{\top}$.
	
	Table \ref{tab:tab2} shows that under our choices the asymptotic test has better performance by far. This suggests that the latter three simulation based misspecification tests may depend drastically on the data generating processes, and choices such as the number of replications and the functions used to calculate the distances. As expected, the asymptotic misspecification test improves as the sample size increases.
	
	Table \ref{tab:tab1} displays average computational time. In particular, the predictive p-value and GoF test based on simulation do not increase their computational time with the sample size, and the former is again the least computational time demanding. On the other hand, the asymptotic and discrepancy tests increase their computational time with the sample size at decreasing rates. .
	
	\subsection{A nonlinear ecological dynamic system}
	We simulate the Rickier prototypical ecological model describing the time evolution of a population $N_{t+1}=rN_t\exp(-N_t+u_i)$ where $r$ is the population growth rate, and it is assumed that the random noise is $u_t\sim N(0, \sigma^2)$. Researchers observe realizations of a Poisson process $Y_t\sim P(\phi N_t)$ \citep{wood2010statistical}.
	
	Let us the data generating process be heteroskedastic, $$u_t\sim N(0, \sigma^2_t), \ 
	\sigma^2_t=\begin{cases}
		\sigma_1^2&\text{ if }1\le t\le t_1\\
		\sigma_2^2&\text{ if }t_1< t\le T
	\end{cases},$$
	where $t_1=\lceil kT \rceil$ and $k\in(0,1]$, $\sigma_1=\sigma_2$ or $k\rightarrow 0$ or $k\rightarrow 1$ imply no misspecification. Observe that the marginal density $f( Y)=\int f( Y, u) \text{d}u$ is analytically intractable.
	
	We set $N_1= 1$, $r=44.7$,  $\phi=10$, $\sigma_1 = 1.3$ and $\sigma_2=0.3$ generating different degrees of misspecification across a grid of values for $k$ over $(0.6,1)$, $n=(250, 500, 1000)$, and 50 simulations.
	
	We set as prior distributions $r\sim \mathcal{U}(40, 70)$, $\phi\sim \mathcal{U}(5, 30)$ and $\sigma\sim \mathcal{U}(0.1, 2)$ using 500,000 draws and setting $\alpha=0.00025$. Summary statistics are the autocorrelations up to 5 of $y_t$, the ordinary least squares estimates of $\beta_1$ and $\beta_2$ from the regression $y_t^{0.3}=\beta_1 y_{t-1}^{0.3}+\beta_2 y_{t-1}^{0.6}+\epsilon_t$, the mean of $y_t$ and number of observed zeros (Wood, 2010).
	
	Algorithm \ref{ABC0} sets $N_n=200,000\geq C\log(n){n^{3/2}}$ and uses 200 bootstrap samples to estimate $\hat{ V}_0$. Algorithm 3 is based on the number of observed zeros, and Algorithm 4 uses $N=10,000$, $\alpha=0.01$ and $h =(\theta^{2\top} \ \theta^{3\top})^{\top}$, ${\theta}=(r \ \phi \ \sigma)^{\top}$. Algorithms 2, 3 and 4 use $R=100$. 
	
	We can see in Table \ref{tab:tab1} that the most computational time demanding is the discrepancy test, whereas the least demanding is the predictive p-value test. Moreover, the computational times of the asymptotic GoF and the discrepancy tests increase with sample size at decreasing rates.   
	
	We can see in Table \ref{tab:tab2} the power results; the power of the GoF based on simulations and the predictive p-value are approximately equal to 0, there is a slightly better performance using the discrepancy test, and the best performance is achieved by the asymptotic GoF test. 
	
	\section{Application: Exchange rate returns}\label{sec: app}
	
	We follow \cite{drovandi2011likelihood} modeling financial returns using g-and-k distributions. We use exchange rate log daily returns from USD/EUR, USD/GBP and GBP/EUR one year before and after the WHO declared the COVID-19 pandemic on 11 March 2020.
	
	\cite{drovandi2011likelihood} propose a moving average of order one using a g-and-k distribution, $z_t=\epsilon_t+\theta_1\epsilon_{t-1}, \ (t=1,\dots,524)$, $\epsilon_t{\sim} N(0,1)$, and $z_t$ is divided by $(1+\theta_1^2)^{1/2}$ to ensure that it marginally has a standard normal distribution. 
	
	We use twelve summary statistics: the seven octiles, the inter-quartile range, robust measures of skewness and kurtosis, and the autocorrelations of order one and two. We perform Algorithm \ref{ABC0} using one million prior draws with an acceptation rate equal to 0.01\%. Table \ref{tab:tab3} in Appendix shows the posterior estimates.
	
	We set $R=100$, and the auxiliary functions are the autocorrelation of order one and $(\theta^{2\top} \ \theta^{3\top})^{\top}$, $\theta=(\theta_1 \  a \ b \ g \ k)^{\top}$ for the predictive p-value and the discrepancy diagnostic, respectively. The asymptotic GoF is based on one million draws, the covariance matrix is based on bootstrap using 200 re-samplings, and the asymptotic distribution is $\chi^2_{(7)}$.
	
	All diagnostics based on the re-sampling distributions do not reject the null hypothesis of no misspecification in all three data sets. On the other hand, the diagnostic based on the asymptotic distribution always rejects the null hypothesis.
	
	\bibliographystyle{apalike}
	\bibliography{refs_mispec}
	
	\clearpage
	\appendix
	\section{Appendix}\label{sec: app}
	\subsection{Maintained assumptions and proofs of main results}\label{ProofT1}
	\subsubsection{Maintained assumptions}
	We first present the relevant assumptions in \cite{frazier2018asymptotic} that we make use of in our results. A useful discussion of these assumptions is provided in Remarks 1-3 of \cite{frazier2018asymptotic}, and for the sake of brevity we do not reinterpret their remarks.

	\begin{assumption}\label{[A1]}
		There exist a non-random map ${b}:{\Theta }%
		\rightarrow \mathcal{B}$, and a sequence of functions $\rho _{n}(u)$ that are monotone non-increasing in $u$ for any $n$ and satisfy $\rho _{n}(u)\rightarrow 0$ as $%
		n\rightarrow \infty $. For fixed $u$, and for all $\theta\in\Theta$,
		\begin{equation*}
			P_{{\theta }}\left[ d\{{\eta }({z}),{b}({\theta })\}>u\right] \leq c({%
				\theta })\rho _{n}(u),\quad \int_{\Theta }c({\theta })d\Pi ({\theta }%
			)<\infty,
		\end{equation*}%
		with either of the following assumptions on $c(\cdot )$:
		
		\noindent{(i)} there exist $c_{0}<\infty $ and $\delta >0$ such that
		for all ${\theta }$ satisfying $d\{{b}({\theta }%
		),{b}({\theta}_{0})\}\leq \delta $ then $c({\theta })\leq c_{0}$;
		
		\noindent{(ii)} there exists $a>0$ such that $\int_{\Theta }c({\theta }%
		)^{1+a}d\Pi ({\theta })<\infty .$
	\end{assumption}

	\begin{assumption} \label{[A2]} There exists some $D>0$ such that, for all $\xi >0$
		and some $C>0$, the prior probability satisfies
		$\Pi \left[ d\{{b}({\theta }),{b}({\theta }_{0})\}\leq \xi \right]
		\geq C \xi ^{D}.$
	\end{assumption}
	
	\begin{assumption}\label{[A3]} {(i)} The map ${b}$
		is continuous. {(ii)} The map ${b}$
		is injective and satisfies:
		$\Vert {\theta }-{\theta }_{0}\Vert \leq L\Vert {b}({\theta })-{b}({\theta }%
		_{0})\Vert ^{\alpha }$ on some open neighbourhood of ${\theta }_{0}$ with $L>0$ and $\alpha >0$. In particular, under the null hypothesis, there exists some $\theta_0\in\Theta$ such that $b(\theta_0)=b_0$. 
	\end{assumption}
	
	\begin{assumption}\label{[A1']}
		There exists a sequence of positive definite matrices $\{\Sigma _{n}({\theta }%
		_{0})\}_{n\ge 1}$, $c_0>0$, $\kappa >1$ and $\delta >0$ such that for all $\Vert {\theta }-{%
			\theta }_{0}\Vert \leq \delta $, $P_{{\theta }}\left[ \Vert \Sigma _{n}({%
			\theta }_{0})\{{\eta }({z})-{b}({\theta })\}\Vert >u\right] \leq {c_{0}%
		}{u^{-\kappa }}$ for all $0<u\leq \delta \|\Sigma_n(\theta_0)\|_*$, uniformly in $n$, for $\|\cdot\|_*$ some matrix norm.
	\end{assumption}
	
	\begin{assumption}\label{[A3']} Assumption \ref{[A3]} holds. The
		map $\theta\mapsto{b}(\theta)$ is continuously differentiable at ${\theta
			_{0}}$ and the Jacobian $\nabla _{\theta }b({\theta }_{0})$ has full column
		rank $k_{\theta }$.
	\end{assumption}
	
	\begin{assumption}
		\label{[A4]} The value
		$\theta_0$ is in the interior of $\Theta$. For some $\delta >0$  and for all $\Vert {%
			\theta }-{\theta }_{0}\Vert \leq \delta $, there exists a sequence of $(k_{\eta }\times k_{\eta
		}) $ positive definite matrices $\{\Sigma_{n}({\theta })\}_{n\ge 1}$, with $k_{\eta
		}=\dim \{\eta (z)\}$, such that for all open sets $B$
		\begin{equation*}
			\sup_{\|\theta - \theta_0\|\leq \delta} \left| P_\theta\left[ {\Sigma }_{n}({\theta })\{{\eta }({z})-{b}({\theta })\} \in B\right] - P\left\{ \mathcal{N%
			}(0,I_{k_{\eta }}) \in B\right\} \right| \rightarrow 0
		\end{equation*} in distribution as $n\rightarrow\infty$, where $I_{k_{\eta }}$ is the $(k_{\eta }\times k_{\eta })$ identity
		matrix.
	\end{assumption}
	
	\begin{assumption}
		\label{[A5]} For all $\Vert \theta -\theta _{0}\Vert \leq \delta $%
		, the sequence of functions ${\theta }\mapsto n^{-1/2}{\Sigma }_{n}({\theta }%
		)$ converges to some positive definite $A(\theta )$ and is
		equicontinuous at ${\theta }_{0}$. \smallskip
	\end{assumption}
	
	\begin{assumption}\label{[A7]} The prior density $\pi(\theta)$ is such that
		{(i)} for $\theta _{0}$ in the interior of ${\Theta }$, $%
		\pi(\theta _{0})>0$; {(ii)} the density function 
		$\pi(\theta)$ is $\beta $- H\"{o}lder in a neighbourhood of $\theta _{0}$.
		{(iii)} For $\Theta \subset \mathbb{R}$, $\int_{\Theta
		}\|\theta \|^{\beta }\pi(\theta )d\theta <\infty $.
	\end{assumption}
	
	
	\subsubsection{Proofs}
	
	We now prove the stated results in Section 3 of the main text. 
	
	\begin{proof}[of Theorem 1] Consider the decomposition
		\begin{align*}
			n^{1/2} V_{0}^{-1/2}\{\hat{\eta}( z)-\eta( y)\}&=n^{1/2} V_{0}^{-1/2}\{\hat{\eta}( z)- b(\hat{{\theta}})\}\\
			&+n^{1/2} V_{0}^{-1/2}\{ b(\hat{{\theta}})- b({\theta}_0)\}-n^{1/2} V_{0}^{-1/2}\{\eta( y)- b({\theta}_0)\}.
		\end{align*}Define $ Z_{n}= n^{1/2}\{\hat{\eta}( z)- b(\hat{{\theta}})\}$ and $ Z_{n}^{0}= n^{1/2}\{{\eta}( y)- b({{\theta}}_0)\}$. By the mean value theorem,
		\begin{flalign}\label{eq:decomp}
			n^{1/2} V_{0}^{-1/2}\{\hat{\eta}( z)-\eta(y)\}&= V_{0}^{-1/2} Z_{n}- V_{0}^{-1/2} Z_{n}^{0}+ V_{0}^{-1/2}\nabla_{{\theta}} b(\bar{{\theta}})n^{1/2}(\hat{{\theta}}-{\theta}_0),
		\end{flalign}where $\bar{{\theta}}$ is a coordinate specific intermediate value.
		
		Theorem 4 in Frazier et~al. (2018), which is valid under the Assumptions \ref{[A1]}-\ref{[A7]}, shows that 
		\begin{equation}
			\label{eq:one}
			\nabla_{{\theta}} b({\theta}_0)n^{1/2}(\hat{{\theta}}-{\theta}_0)=\Sigma^{-1} B_0^{\intercal} V_{0}^{-1} Z_{n}^{0}+o_{P}(1),
		\end{equation}
		where $o_{P}(1)$ denotes a random variable that converges to zero in probability, and where $B_0=\nabla_{{\theta}} b({\theta}_0)$ and $\Sigma= B_0^{\intercal} V_{0}^{-1} B_0$. Write $X=V_{0}^{-1/2} B_0$, and apply equation \eqref{eq:one} into equation \eqref{eq:decomp}, and re-arrange terms to obtain
		\begin{flalign}
			n^{1/2} V_{0}^{-1/2}\{\hat{\eta}( z)-\eta(y)\}&= V_{0}^{-1/2} Z_{n}- V_{0}^{-1/2} Z_{n}^{0}\nonumber\\
			&+ V_{0}^{-1/2}B_0\Sigma^{-1}B_0^\intercal V_0^{-1/2}V_{0}^{-1/2}Z_n^0+o_P(\|n^{1/2}\{\hat\theta-\theta_0\}\|)\nonumber\\
			&=V_{0}^{-1/2} Z_{n}- (I- P_{X})V_{0}^{-1/2}Z_{n}^{0}+o_P(1)\label{eq:proj},
		\end{flalign}where we have used that fact that $\nabla_\theta b(\theta)$ is continuous on $\|\theta-\theta_0\|\le\delta$, for some $\delta$, and $\|\bar\theta-\theta_0\|=o_P(1)$, since $\hat\theta$ is consistent for $\theta_0$ (see Theorem 4 in \cite{frazier2018asymptotic} for details).
		
		Now, we show that $Z_n=o_P(1)$ under our particular simulation design. Write $$ Z_{n}=\frac{n^{1/2}}{N_{n}^{1/2}}\left[\frac{1}{N_{n}^{1/2}}\sum_{i=1}^{N_{n}}\left\{\hat{\eta}({ z}_{i})- b(\hat{{\theta}})\right\}\right].$$ By Assumption \ref{[A4]}, $$ Z_{n}=\frac{n^{1/2}}{N_{n}^{1/2}}O_{P}(1).$$ Therefore, $ Z_{n}=o_{P}(1)$ provided $n^{1/2}/N_{n}^{1/2}=o(1)$. Our chosen simulation design has explicitly taken $N_n\ge C \ln(n){n}^{q/2}$, so that, for $q=\max\{k_\theta,2\}$,  $$n^{1/2}/N_{n}^{1/2}\le C n^{1/2}/(\ln(n){n}^{q/2})^{1/2} \le C(\ln(n))^{-1/2}n^{1-q}=o(1).$$Hence,
		\begin{equation}\label{eq:Zn}
			Z_{n}=o_{}(1)O_{P}(1)= o_{P}(1)
		\end{equation}as desired.
		
		Applying equation \eqref{eq:Zn} into equation \eqref{eq:decomp} yields
		\begin{flalign}\label{eq:tog}
			n^{1/2} V_{0}^{-1/2}\{\hat{\eta}( z)-\eta( y)\}&=o_{P}(1)-(I- P_{ X}) V_{0}^{-1/2} Z_{n}^{0}.
		\end{flalign} Write $$\hat{V}_0^{-1}=V_0^{-1/2}V_0^{1/2}\hat{V}_0^{-1}V_0^{1/2}V_0^{-1/2}.$$ Apply the above and equation \eqref{eq:tog} to obtain
		\begin{align*}
			\mathcal{J}_n\{\hat{ \eta}( z), \eta( y)\}&=n^{1/2}\left\{\hat{ \eta}( z)- \eta( y)\right\}^{\intercal} V_0^{-1/2}V_0^{1/2}\hat{V}_0^{-1}V_0^{1/2}V_0^{-1/2}n^{1/2}\left\{\hat{ \eta}( z)- \eta( y\right\}\\&=n^{1/2}\left\{\hat{ \eta}( z)- \eta( y)\right\}^{\intercal} V_0^{-1/2}V_0^{-1/2}n^{1/2}\left\{\hat{ \eta}( z)- \eta( y\right\}+o_P(1)\\
			&=\left\{(I- P_{ X}) V_{0}^{-1/2} Z_{n}^{0}\right\}^{\intercal}\left\{(I- P_{ X}) V_{0}^{-1/2} Z_{n}^{0}\right\}+o_{P}(1),    
		\end{align*} where the second equality follows since $\hat{V}_0^{-1}$ is a consistent estimator of $V_0^{-1}$, and the last equality by applying \eqref{eq:tog}. The stated result follows by the asymptotic normality of $ Z_n^0$, Assumption \ref{[A4]}, and the fact that $\text{Rank}\left( I-P_{ X}\right)=k_{\eta}-k_{{\theta}}$.

	\end{proof}

	
	\begin{proof}[of  Corollary 1]
		First observe that, by Assumption \ref{[A1]}, $d\{ \eta( y), \eta( z)\}$ converges to $d\left\{ b_0,b(\theta)\right\}$ uniformly in $\theta$, where we recall that $b_0=\plim_n\eta(y)$. Define, 
		$$
		\theta^\star=\arg\min_{\theta\in\Theta}d\{ b_0, b(\theta)\}.
		$$From Corollary 1 in \cite{frazier2020model}, and the assumptions of the result, $\hat\theta=\theta^\star+o_P(1)$. Thus, from Assumption 1 and consistency of $\hat{V}_0$, we see that 
		$$
		\mathcal{J}\{\hat{\eta}( z),\eta( y)\}/n=\{ b(\theta^\star)- b_0\}^{\intercal} V_{0}^{-1}\{ b(\theta^\star)- b_0\}+o_P(1).$$ Moreover, since $V_0$ is positive-definite, we have that 
		$$
		\mathcal{J}\{\hat{\eta}( z),\eta( y)\}/n=\{ b(\theta^\star)- b_0\}^{\intercal} V_{0}^{-1}\{ b(\theta^\star)- b_0\}+o_P(1)\ge C\|b(\theta^\star)-b_0\|^2\{1+o_P(1)\},
		$$for some $C>0$. Since all norms are equivalent in Euclidean space, $d\left\{ b_0,b(\theta^\star)\right\}>0$ implies that $\|b_0-b(\theta^\star)\|^2>0$. 
		
		Consequently, 
		$$
		\mathcal{J}\{\hat{\eta}( z),\eta( y)\}\ge n\cdot C\|b(\theta^\star)-b_0\|^2\{1+o_P(1)\}>0,
		$$
		and the right-hand side of the above diverges as $n$ diverges. 
	\end{proof}
	
	\subsection{Regression correction in approximate Bayesian computation}
	
	Intuitively, this can be seen by simply considering the linear regression adjustment approach under misspecification. An regression correction in approximate Bayesian computation approach that first runs Algorithm 1 in main paper, with tolerance $\epsilon_{n}(\alpha)$, to obtain a set of selected draws and summaries $\{{\theta}^i, \eta({ z}^i)\}$, and then uses a linear regression model to predict the values of ${\theta}$ generated from the approximate Bayesian computation measure $\Pi_{\epsilon}\left\{\cdot\mid\eta({ y})\right\}$. An accepted value from Algorithm 1 in main paper, ${\theta}^i$, is then artificially related to $\eta({ y})$ and $\eta({ z})$ through the linear regression model 
	$${\theta}^i=\mu+\beta^{\top}\left\{\eta({ y})-\eta({ z}^i)\right\}+\nu_i,$$ 
	where $\nu_i$ is the model stochastic error. Define $\hat{{\theta}}=\sum_{i=1}^{\delta}{\theta}^i/\delta$ and $\hat{\eta}( z)=\sum_{i=1}^{\delta}\eta({ z}^i)/\delta$. Regression correction in approximate Bayesian computation defines the adjusted parameter draw according to
	\begin{align}
		\tilde{{\theta}}^i&={\theta}^i+\hat{\beta}^{\top}\left\{\eta({ y})-\eta({ z}^{i})\right\}, \label{eq4}\\
		\hat{\beta}&= \left[\frac{1}{\delta}\sum_{i=1}^{\delta}\left\{ \eta({ z}^{i})-\hat{ \eta}( z)\right\}\left\{\eta({ z}^{i})-\hat{\eta}( z)\right\}^{\top}\right]^{-1}\left[\frac{1}{\delta}\sum_{i=1}^{\delta}\left\{\eta({ z}^{i})-\hat{\eta}( z)\right\}\left({\theta}^i-\hat{{\theta}}\right)\right].\label{eq5}
	\end{align}
	For ${{\theta}}^i\sim \Pi_{\epsilon}\left\{d{\theta}\mid\eta({ y})\right\}$, the value $\tilde{{\theta}}^{i}$ is noting but a scaled and shifted version of ${\theta}^i$. More importantly, this scaling and shifting is related to $\eta( y)-\eta( z)$. Therefore, if $d\{\eta( y),\eta( z)\}$ is large, as is the case when the model is misspecified, it is likely that $\hat{\beta}^{\top}\{ \eta( y)-\eta( z)\}$ will be large, in absolute value, and all the regression adjustment will force $\tilde{{\theta}}^{i}$ away from ${\theta}^{i}$.

	\subsection{Algorithms}\label{sec:alg}
	
	\begin{algorithm}
		\caption{Algorithm 1: Goodness of fit based on the asymptotic distribution}\label{ABC1}
		\begin{tabbing}	
			\enspace (1) Given the results from Algorithm 1 in main paper\\
			
			\enspace (2) Simulate $\hat{ z}$, a pseudo-data set of size $N_n$ from the model $p(\cdot\mid\hat{ \theta})$\\
			
			\enspace (3) Calculate $\hat{\eta}( z)$, $\hat{ V}_0$ and $DT_n$\\
			
			\enspace (4) Reject the null hypothesis of no misspecification if $DT_n > \chi^2_{k_{\eta}-k_{\theta}}$\\		 
		\end{tabbing}
		\vspace*{-25pt}
	\end{algorithm}
	
	\begin{algorithm}
		\caption{Algorithm 2: Goodness of fit based on the simulated distribution}\label{ABC2}
		\begin{tabbing}	
			\enspace (1) Given the results from Algorithm 1 in main paper\\
			
			\enspace (2) Calculate the average Euclidean distance $\bar{\epsilon}=\frac{1}{N}\sum_{i=1}^{N} d\left\{\eta( y),\eta( z^i)\right\}$\\
			
			\enspace (3) {\texttt{$r=1,\dots,R$}}\\
			
			\qquad (i) Sample $\eta( z_r)$ without replacement from $\eta( z_i), i = 1,\dots,N$\\ 
			
			\qquad (ii) Perform regression correction in approximate Bayesian computation \\
			using $\eta( z_r)$
			as summary statistics from the pseudo-data $ z_r$\\ 
			
			\qquad (iii) Calculate the average Euclidean distance $\bar{\epsilon}_r=\frac{1}{N}\sum_{i=1}^{N} d\left\{\eta( z_r),\eta( z_i)\right\}$\\  
			
			\enspace (4) Order $\bar{\epsilon}_r$ in increasing order $\bar{\epsilon}_{(1)}\leq \bar{\epsilon}_{(2)}\leq\dots\leq\bar{\epsilon}_{(R)}$\\
			
			\enspace (5) Do not reject the null hypothesis of no misspecification if $\bar{\epsilon}\leq \bar{\epsilon}^{(1-\alpha)}$, \\
			where $\bar{\epsilon}^{(1-\alpha)}$ is
			the $1-\alpha$ quantile of $\bar{\epsilon}_r$\\
		\end{tabbing}
		\vspace*{-25pt}
	\end{algorithm}

	\begin{algorithm}
		\caption{Algorithm 3: Predictive p-value}\label{ABC3}
		\begin{tabbing}	
			\enspace (1) Given the results from Algorithm 1 in main paper\\
			
			\enspace (2) {\texttt{$r=1,\dots,R$}}\\
			
			\qquad (i) Sample $ \theta_r$ with replacement from the $\delta$ draws of $\Pi_{\epsilon}\left\{\cdot\mid\eta( y)\right\}$\\ 
			
			\qquad (ii) Simulate $\hat{ z}_r$, a pseudo-data set of size $n$ from the model $p(\cdot\mid \theta_r)$\\ 
			
			\qquad (iii) Calculate $\eta(\hat{ z}_r)$, where $\eta(\hat{ z}_r)$ is a one dimensional statistic used to perform\\
			the marginal predictive analysis\\  
			
			\enspace (3) Order $\eta(\hat{ z}_r)$ in increasing order $\eta(\hat{ z})_{(1)}\leq \eta(\hat{ z})_{(2)}\leq\dots\leq\eta(\hat{ z})_{(R)}$\\
			
			\enspace (4) Do not reject the null hypothesis of no misspecification if \\
			$\eta(\hat{ z})^{(\alpha/2)} \leq \eta( y)\leq \eta(\hat{ z})^{(1-\alpha/2)}$, where $\eta(\hat{ z})^{(\alpha/2)}$ and $\eta(\hat{ z})^{(1-\alpha/2)}$\\
			are the $\alpha/2$ and $1-\alpha/2$ quantiles of $\eta(\hat{ z}_r)$ \\
		\end{tabbing}
		\vspace*{-25pt}
	\end{algorithm}

	\begin{algorithm}
		\caption{Algorithm 4: Discrepancy diagnostic}\label{ABC4}
		\begin{tabbing}	
			\enspace (1) Given the results from Algorithm 1 in main paper\\
			
			\enspace (2) Implement the regression correction approximate Bayesian\\
			computation algorithm
			using equations \ref{eq4} and \ref{eq5}\\
			
			\enspace (3) Calculate $d=n^{1/2}\mid\mid\hat{h}-\tilde{h}\mid\mid$, where $\mid\mid\cdot\mid\mid$ is the Euclidean distance, $\hat{h}$ and $\tilde{h}$\\ are average functions of $ \theta$ calculate from the posterior draws of accept and reject, and 
			\\regression correction
			approximate Bayesian computation, respectively\\
			
			\enspace (4) {\texttt{$r=1,\dots,R$}}\\
			
			\qquad (i) Simulate $\hat{ z}_r$, a pseudo-data set of size $n$ from the model $p(\cdot\mid\hat{ \theta})$\\ 
			
			\qquad (ii) Perform accept and reject, and regression correction approximate Bayesian\\ computation using $\eta(\hat{ z}_r)$ as
			summary statistics from the pseudo-data $\hat{ z}_r$\\ 
			
			\qquad (iii) Calculate $d_r=n^{1/2}\mid\mid\hat{h}_r-\tilde{h}_r\mid\mid$\\ 
			
			\enspace (5) Order $d_r$ in increasing order $d_{(1)}\leq d_{(2)}\leq\dots\leq d_{(R)}$\\
			
			\enspace (6)  Do not reject the null hypothesis of no misspecification if $d\leq d_r^{(1-\alpha)}$,\\
			where $d_r^{(1-\alpha)}$
			is the $1-\alpha$ quantile of $d_r$\\
		\end{tabbing}
		\vspace*{-25pt}
	\end{algorithm}
	\clearpage
	\subsection{Application: Exchange rate returns}
	
	\begin{table}[htbp]\centering \caption{Posterior estimates of the log return models\label{tab:tab3}}
		\begin{threeparttable}
			\resizebox{1\textwidth}{!}{\begin{minipage}{\textwidth}
					\begin{tabular}{c c c c}
						Parameter	&	USD/EUR	&	USD/GBP	&	GBP/EUR\\
						\multirow{2}{*}{$\theta_1$}	&	0.208	& 0.292		&	-0.158	\\
						&	(-0.898, 0.943)	&	(-0.935, 0.958)	& (-0.969, 0.799)		\\
						\multirow{2}{*}{$a$}	&	0.076	&	0.084	&	 0.076	\\
						&	(0.006, 0.208)	& (0.009, 0.194)	& (0.003, 0.209)		\\
						\multirow{2}{*}{$b$}	&	0.082	&	0.075	&	0.082	\\
						&	(0.007, 0.248)	&	(0.003, 0.247)	& (0.007, 0.264)		\\
						\multirow{2}{*}{$g$}	&	-0.215	&	-0.215	&	-0.027	\\
						&	(-3.065, 2.448)	&	(-3.398, 1.837)	& (-2.788, 2.607)		\\
						\multirow{2}{*}{$k$}	&	0.018	&	0.046	&	0.018	\\
						&	(-0.447, 0.697)	&	(-0.427, 0.655)	& (-0.447, 0.669)		\\
						
					\end{tabular}
					\begin{tablenotes}[para,flushleft]
						\footnotesize \textit{Notes}: Median upper, and 95\% symmetric credible intervals in parenthesis. The 95\% credible intervals of the autocorrelation, skewness and kurtosis coefficients do embrace zero in the three data sets.
						
					\end{tablenotes}
			\end{minipage}}
		\end{threeparttable}
	\end{table}
	
\end{document}